\documentclass[aps, pra, reprint, onecolumn, showpacs, amsmath, amssymb]{revtex4-1}

\usepackage{bm}
\usepackage{amsmath, amscd, amssymb}
\usepackage{graphicx}
\usepackage{enumerate}
\usepackage{subcaption}
\usepackage{gensymb}
\usepackage[english]{babel}
\usepackage{soul}
\usepackage[usenames]{color}

\DeclareMathOperator{\sgn}{\textrm{sgn}}

\begin{document}
\title{ Radiation of charge moving through a dielectric spherical target: \\ ray optics and aperture methods }

\author{ Andrey V. Tyukhtin }
\email{a.tyuhtin@spbu.ru}
\author{ Ekaterina S. Belonogaya }
\author{ Sergey N. Galyamin }
\author{ Victor V. Vorobev }
\affiliation{ Saint Petersburg State University, 7/9 Universitetskaya nab., St. Petersburg, 199034 Russia }

\date{\today}

\begin{abstract}
Radiation of charged particles moving in the presence of dielectric targets is of significant interest for various applications in the accelerator and beam physics. 
The size of these targets is typically much larger than the wavelengths under consideration. 
This fact gives us an obvious small parameter of the problem and allows developing approximate methods for analysis. We develop two methods, which are called the ``ray optics method'' and the ``aperture method''. 
In the present paper, we apply these methods to analysis of Cherenkov radiation from a charge moving through a vacuum channel in a solid dielectric sphere. 
We present the main analytical results and describe the physical effects. In particular, it is shown that the radiation field possesses an expressed maximum at a certain distance from the sphere at the Cherenkov angle. 
Additionally, we perform simulations in COMSOL Multiphysics and show a good agreement between numerical and analytical results. 
\end{abstract}

\maketitle


\section{\label{sec:1}Introduction}

Radiation of charged particles moving in the presence of dielectric objects (``targets'') is of vital interest for various applications~\cite{b:1,b:2,b:3,b:4,b:5}. For example, several experiments have shown that prismatic and conical targets can be prospective for both bunch diagnostics and generation of high-power radiation~\cite{b:2,b:3,b:4,b:5}. Further development of these topics requires an accurate calculation of Cherenkov radiation (CR) outside dielectric objects, which is typically impossible to do rigorously due to the complicated geometry of these objects. The only exceptions allowing the construction of rigorous solutions are the simplest geometries like infinite cylinder~\cite{b:6,b:7,b:8,b:9} or sphere~\cite{b:9,b:10}. However, the obtained formulas (infinite series) can be applied to the field analysis only when the wavelength $\lambda$ is comparable with the target radius. Moreover, such ``practical'' modifications of the geometry as a vacuum channel for the charge flight cannot be incorporated in this solution. 

For the discussed applications, the vacuum channel is needed for the flight of the bunch. Moreover, the target dimensions are typically much larger than the wavelengths of interest. Therefore, both calculations based on the solutions mentioned above and numerical simulations are very complicated. However, the mentioned relation between $\lambda $ and target size gives us an obvious small parameter allowing the development of approximate (asymptotic) methods for the analysis of radiation. Recently we have offered and successfully verified two such methods called the ``ray optics method''~\cite{b:11,b:12} and ``aperture method''~\cite{b:13,b:14,b:15,b:16,b:17,b:18,b:19}. They can be divided into three steps.

The first two steps are the same for both methods. First, we solve the specific ``etalon'' problem, which does not take into account the ``external'' boundaries of the target. For example, if the charge moves in the vacuum channel inside the target, then in the first step, we consider the problem with the channel in the unbounded medium. In other words, we consider only the boundary nearest to the charge trajectory and solve the problem for the semi-infinite medium. 

In the second step, we select a part of the external surface of the object which is illuminated by CR and transparent for CR (so that there is no total internal reflection here). This part of the object boundary is called an ``aperture'' further. Then we use the fact that the object is large in comparison with the wavelengths under consideration. More precisely, we assume that: (i) the size of the aperture $\Sigma $ is much larger than the wavelength $\lambda $; (ii) the distance from the main part of the aperture to the charge trajectory is also much larger than the wavelength $\lambda$.

The field obtained in the first step is used as the incident field on the aperture. Due to point (ii), we can neglect the quasi-static (quasi-Coulomb) part of this field and use corresponding asymptotic approximation, which is a quasi-plane wave (more precisely, a cylindrical wave with a small curvature of the wave front). Further, we decompose this wave into a superposition of vertical and horizontal polarizations (concerning the plane of incidence) and calculate the field on the external surface of the aperture using Snell’s law and the Fresnel equations. 

The third step is different for the two methods. The ray optics method uses the ray optics laws for calculation of the wavefield outside the object~\cite{b:11,b:12}. However, this technique has essential limitations. First, the so-called ``wave parameter'' $W \sim \lambda L/\Sigma $ ($L$ is a distance from the aperture to the observation point) should be small $W \ll 1$. Note that this means that the distance $L$ can be much larger than $\lambda $ but cannot be larger than $\Sigma /\lambda $. In particular, we cannot consider the important Fraunhofer area where $W \gg 1$. Second, the observation point can not be in the neighborhood of focuses and caustics, where ray optics is not applicable. 

The aperture method is more general~\cite{b:13,b:14,b:15,b:16,b:17,b:18,b:19} than the ray optics. It is valid for observation points with arbitrary wave parameter $W$, including the Fraunhofer (far-field) area and neighborhoods of focuses and caustics. In the third step of this technique, we calculate the field outside the target using Stratton-Chu formulas (``aperture integrals''). These formulas allow determining the field in the surrounding space if tangential components of the electric and magnetic fields on the aperture are known. 

This paper is devoted to the study of CR from a solid dielectric sphere with the radius being much larger than the wavelength and having an axisymmetric vacuum channel where the charge moves. Such a target can be manufactured with high accuracy and can be a prospective candidate for the aforementioned applications. We apply both methods for the spherical target and compare the obtained analytical results with the results of COMSOL Multiphysics simulations. 

\section{\label{sec:2}The field on the ball surface}

\begin{figure*}[t!]
    \centering
    \includegraphics[scale=0.8]{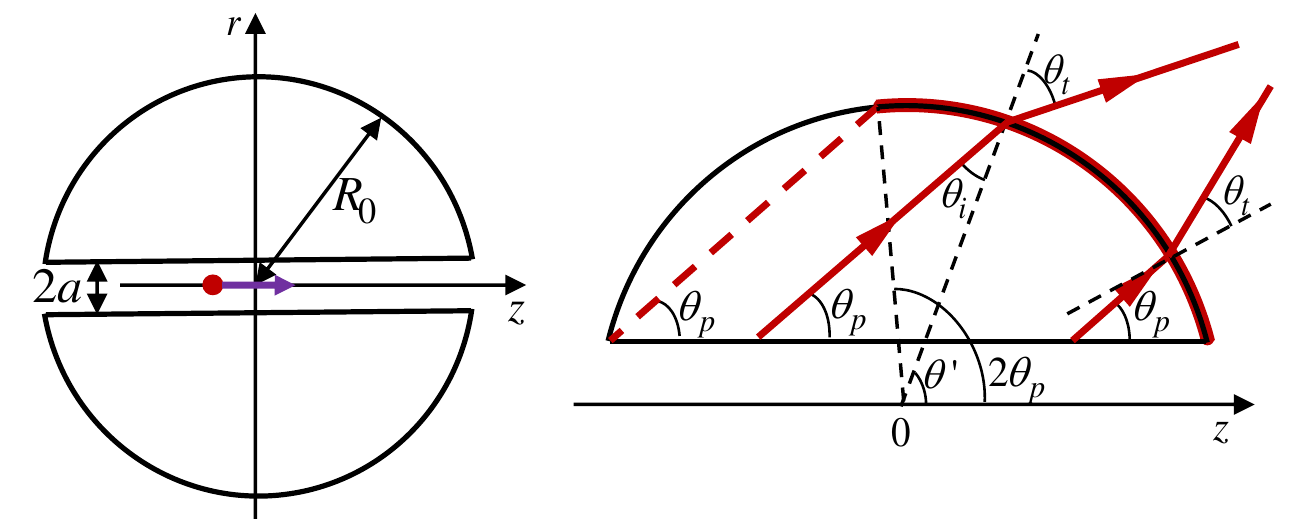}
    \caption{Cross-section of the dielectric ball with vacuum channel (left); the incident and refracted waves (right). Incidence angle ${{\theta}_{i}}$ and refraction angle ${{\theta}_{t}}$ are positive for the left ray, and negative for the right ray; the illuminated part of the sphere (aperture) is highlighted by the bold red line.}\label{fig:1}
\end{figure*}

Here we consider a dielectric ball with the radius $R_0$ having the cylindrical vacuum channel with the radius $a$ (Fig.~\ref{fig:1}, left). In accordance with point (ii) in the Introduction, it is assumed that $kR_0 \gg 1$, where $k=\omega /c$ ($\omega $ is the frequency, $c$ is the speed of light in vacuum). The ball material is characterized by permittivity $\varepsilon $, permeability $\mu $, and the refractive index $n=\sqrt{\varepsilon \mu }$ (the conductivity is considered to be negligible). The channel axis ($z$-axis) coincides with the ball diameter. The charge $q$ moves with constant velocity $\vec{V}=c\beta {{\vec{e}}_{z}}$ along the $z$-axis, and this velocity exceeds the ``Cherenkov threshold'', i.e. $\beta >{1}/{n}$.

For definiteness, we deal with a point charge having the charge density $\rho =q\delta (x)\delta (y)\delta (z-Vt)$ where $\delta (\xi )$ is the Dirac delta function. However, the results obtained further can be easily generalized for the case of a thin bunch with finite length because we consider Fourier transforms of the field components. Further, we use the spherical ($R, \theta, \varphi $) and cylindrical ($r, \varphi, z$) coordinate systems. 

First, we find the ``incident'' field, i.e., solution of the ``etalon'' problem (field in the infinite medium with the vacuum channel). For the case under consideration, this field is well known~\cite{b:21}. We are interested in the incident field on the ball surface at the point $R_0, \theta ', \varphi '$. Considering that $kR_0 \gg 1$ one can write the Fourier-transform of the magnetic component in the form of the cylindrical wave (we use the Gaussian system of units): 
\begin{equation}\label{eq:2.1}
    H_{\varphi '}^{\left( i \right)}\left(R_0, \theta' \right)\approx \frac{q}{c}\eta \sqrt{\frac{s}{2\pi r'}}\exp \left\{ i\left( sr'+\frac{\omega }{V}z'-\frac{\pi }{4} \right) \right\},
\end{equation}
\begin{equation}\label{eq:2.2}
    \eta =-\frac{2i}{\pi a}{{\left[ \kappa \frac{1-{{n}^{2}}{{\beta }^{2}}}{\varepsilon \left( 1-{{\beta }^{2}} \right)}{{I}_{1}}\left( \kappa a \right)H_{0}^{\left( 1 \right)}\left( sa \right)+s{{I}_{0}}\left( \kappa a \right)H_{1}^{\left( 1 \right)}\left( sa \right) \right]}^{-1}},
\end{equation}
where $r'=R_0\sin \theta '$, $z'=R_0\cos \theta '$, $s(\omega )=k{{\beta }^{-1}}\sqrt{{{n}^{2}}{{\beta }^{2}}-1}$, $\kappa (\omega )=k{{\beta }^{-1}}\sqrt{1-{{\beta }^{2}}}$, ${{I}_{0, 1}}(x)$  are the modified Bessel functions, $H_{0, 1}^{(1)}(x)$ are the Hankel functions. Note that $\operatorname{Im}s\left( \omega  \right)\ge 0$ if we take into account a small dissipation. If dissipation tends to zero, then this condition results in the rule $\sgn \left( s\left( \omega  \right) \right)=\sgn \left( \omega  \right)$ (we exclude the exotic case of the so-called ``left-handed'' medium). The result~\eqref{eq:2.1} is valid for $\left| sr' \right| \gg 1$. The electric field ${{\vec{E}}^{(i)}}$ can be easily found because vectors ${{\vec{E}}^{(i)}}$, ${{\vec{H}}^{(i)}}$ and the wave vector of Cherenkov radiation ${{\vec{k}}^{(i)}}=s{{\vec{e}}_{r}}+{{{{\vec{e}}}_{z}}\omega }/{V}$ form the right-hand orthogonal triad in this area, thus ${{\vec{E}}^{(i)}}=-\sqrt{{\mu }/{\varepsilon }}\left[ {{{{\vec{k}}}^{(i)}}}/{{{k}^{(i)}}}\times {{{\vec{H}}}^{(i)}} \right]$. The angle between the wave vector ${{\vec{k}}^{(i)}}$ and the charge velocity $\vec{V}$ is ${{\theta }_{p}}=\arccos \left( {1}/{\left( n\beta  \right)} \right)$.

Applying Snell’s law and the Fresnel equations (note that waves have only vertical polarization) one can obtain the following expressions for the field components on the outer surface of the ball: 
\begin{equation}\label{eq:2.3}
    {{H}_{\varphi '}}\left(R_0, \theta' \right)={{T}_{v}}(\theta ')H_{\varphi '}^{(i)}\left(R_0, \theta' \right), 
    \quad
    {{E}_{\theta '}}\left(R_0, \theta' \right)={{H}_{\varphi '}}\left(R_0, \theta' \right) \cos{{\theta }_{t}}(\theta '),
\end{equation}
where       
\begin{equation}\label{eq:2.4}
    {{T}_{v}}(\theta ')=\frac{2\cos {{\theta }_{i}}(\theta ')}{\cos {{\theta }_{i}}(\theta ')+\sqrt{{\varepsilon }/{\mu }} \cos {{\theta }_{t}}(\theta ')},
\end{equation}
\begin{equation}\label{eq:2.5}
    {{\theta }_{t}}(\theta ')=\arcsin \left( n\sin {{\theta }_{i}}(\theta ') \right),
    \quad
    {{\theta }_{i}}(\theta ')=\theta '-{{\theta }_{p}}.
\end{equation}
Here ${{\theta }_{i}}(\theta ')$ is the angle of incidence of the wave on the surface, ${{\theta }_{t}}(\theta ')$ is the angle of refraction (see Fig.~\ref{fig:1}, right, where rays with positive and negative angles ${{\theta }_{i}}$ and  $ \theta_t $ are shown). 

\section{\label{sec:3}Ray optics method}

According to the ray optics approach, in order to determine the field at the given observation point $r, \varphi, z$, one should first determine the ray which starts at a certain point ${r}'({\theta }')=R_0\sin {\theta }'$, ${\varphi }'$,  ${z}'({\theta }')=R_0\cos {\theta }'$ at the aperture and reaches this observation point. Due to the symmetry over $\varphi $ we obtain $\varphi ={\varphi }'$ and 
\begin{equation}\label{eq:3.1}
    r={r}'(\theta ')+l\sin (\theta '-{{\theta }_{t}}), \quad    z={z}'(\theta ')+l\cos (\theta '-{{\theta }_{t}}),
\end{equation}
where $l$ is the length of the ray. Therefore for each pair $r, z$ corresponding pairs ${\theta }', l$ should be determined from~\eqref{eq:3.1}. This problem can be solved numerically. 

\begin{figure}[t!]
    \centering
    \includegraphics[width=0.5\linewidth]{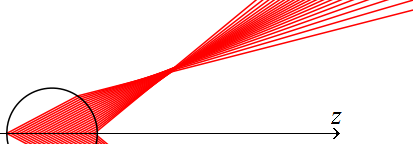}
    \caption{The ray picture for $\varepsilon =2$, $\beta =0.8$.}\label{fig:2}
\end{figure}

A typical example of the rays structure is shown in Fig.~\ref{fig:2}. One can see that rays intersect each other (solution of~\eqref{eq:3.1} is not unique) and form caustics. Moreover, the considerable concentration of the rays occurs near the ray which is not refracted ($\theta '={{\theta }_{p}}$) at some distance from the ball. The area of the concentration of the rays corresponds to the area where the field increases. The field along each ray can be written in the following form (for example, we consider Fourier-transform of $\varphi $-component of the magnetic field): 
\begin{equation}\label{eq:3.2}
    {{H}_{\varphi }}\left(R, \theta \right)={{H}_{\varphi }}\left( R_0,  {{\theta }'} \right)\sqrt{{D\left( 0 \right)}/{D\left( l \right)}}\exp (ikl)
\end{equation}
(the electric field is equal to the magnetic field and orthogonal to it and the ray). Here ${{H}_{\varphi }}\left(R_0, \theta' \right)$ is the corresponding Fourier-transform at the point of the ray exit and $D(l)$ is the square of the ray tube cross-section~\cite{b:20}. Square root in~\eqref{eq:3.2} describes the change in the field magnitude due to the divergence (or convergence) of the ray tube. Using~\eqref{eq:3.1}, Cartesian coordinates of the observation point can be obtained as the function of $\theta'$ and $\varphi'$: $x({\theta }', {\varphi }')=r(\theta ') \cos{\varphi }'$, $y({\theta }', {\varphi }')=r(\theta ') \sin{\varphi }'$.  $D(l)$ can be calculated as follows~\cite{b:20}:
\begin{equation}\label{eq:3.3}
    D\left( l \right)=\frac{1}{\sqrt{g}}\left| \begin{matrix}
   \kappa _{x}^{*} & \kappa _{y}^{*} & \kappa _{z}^{*}  \\
   {\partial x}/{\partial \varphi '} & {\partial y}/{\partial \varphi '} & {\partial z}/{\partial \varphi '}  \\
   {\partial x}/{\partial \theta '} & {\partial y}/{\partial \theta '} & {\partial z}/{\partial \theta '}  \\
\end{matrix} \right|,
\end{equation}
where $g=R_{0}^{4}\cdot {{\sin }^{2}}\theta '$ is a determinant of the metric tensor for the sphere, ${{\vec{\kappa }}^{*}}$ is a unit vector along refracted ray: 
$\kappa _{x}^{*}=\sin (\theta '-{{\theta }_{t}})\cos{\varphi }'$, 
$\kappa _{y}^{*}=\sin (\theta '-{{\theta }_{t}})\sin{\varphi }'$, 
$\kappa _{z}^{*}=\cos (\theta '-{{\theta }_{t}})$.
After a series of transformations the following expression can be obtained: 
\begin{equation}\label{eq:3.5}
    D\left( l \right)=\cos {{\theta }_{t}}-\frac{l}{R_0}\left[ \frac{\sin \left( {{\theta }_{t}}-{{\theta }_{i}} \right)}{\sin {{\theta }_{i}}\cos {{\theta }_{t}}}+\frac{\sin \left( {{\theta }_{t}}-\theta ' \right)\cos {{\theta }_{t}}}{\sin \theta '} \right]+{{\left( \frac{l}{R_0} \right)}^{2}}\frac{\sin \left( {{\theta }_{t}}-{{\theta }_{i}} \right)\sin \left( {{\theta }_{t}}-\theta ' \right)}{\sin {{\theta }_{i}}\cos {{\theta }_{t}}\sin \theta '}.
\end{equation}
Numerical results based on~\eqref{eq:3.2} will be given in Section~\ref{sec:5}. 

\section{Aperture method}
\label{sec:4}

The aperture integrals (Stratton-Chu formulas) for Fourier transform of the electric field can be written in the following general form~%
\cite{b:15, b:16, b:17, b:18, b:19}%
:
\begin{equation}\label{eq:4.1}
    \begin{aligned}
  & \vec{E}\left( {\vec{R}} \right)={{{\vec{E}}}^{(h)}}\left( {\vec{R}} \right)+{{{\vec{E}}}^{(e)}}\left( {\vec{R}} \right), \\ 
 & {{{\vec{E}}}^{(h)}}\left( {\vec{R}} \right)=\frac{ik}{4\pi }\int\limits_{\Sigma }{\left\{ \left[ \vec{n}'\times \vec{H}\left( \vec{R}' \right) \right]G\left( \left| \vec{R}-\vec{R}' \right| \right) \right.}\left. +\frac{1}{{{k}^{2}}}\left( \left[ \vec{n}'\times \vec{H}\left( \vec{R}' \right) \right]\cdot \nabla ' \right)\nabla 'G\left( \left| \vec{R}-\vec{R}' \right| \right) \right\}d\Sigma ', \\ 
 & {{{\vec{E}}}^{(e)}}\left( {\vec{R}} \right)=\frac{1}{4\pi }\int\limits_{\Sigma }{\left[ \left[ \vec{n}'\times \vec{E}\left( \vec{R}' \right) \right]\times \nabla 'G\left( \left| \vec{R}-\vec{R}' \right| \right) \right]d\Sigma ', } \\ 
\end{aligned}
\end{equation}
where $\Sigma $ is the aperture area, $\vec{E}\left( \vec{R}' \right)$, $\vec{H}\left( \vec{R}' \right)$ is the field on the surface of the aperture, the prime sign indicates that operator or coordinate is referred to the surface of an object, $k={\omega }/{c}$, $\vec{n}'$ is the unit external normal to the aperture in the point $\vec{R}'$, $G\left( R \right)={\exp \left( ikR \right)}/{R}$ is the Green function of Helmholtz equation, and ${\nabla }'$ is the gradient: $\nabla '={{\vec{e}}_{x}}{\partial }/{\partial x'}+{{\vec{e}}_{y}}{\partial }/{\partial y'}+{{\vec{e}}_{z}}{\partial }/{\partial z'}$. Analogous formulas are known for the magnetic field as well. Note that $\left| {\vec{E}} \right|\approx \left| {\vec{H}} \right|$ in the region several wavelengths far from the aperture. 

In the case of the spherical object, it is convenient to write aperture integrals using spherical coordinates $R, \theta, \varphi $. Besides the primary condition $kR_0 \gg 1$, we impose for simplicity an additional condition $k(R-R_0) \gg 1$, which means that the observation point is located at a distance of no less than several wavelengths from the ball surface. Using the cylindrical symmetry of the problem, we can choose an observation point on the plane $x, z$ ($\varphi =0$). As a result, one can obtain from~\eqref{eq:4.1} the following expressions for Fourier-transforms of the non-zero electric field components:
\begin{equation}\label{eq:4.2}
    \vec{E}={{\vec{E}}^{(h)}}+{{\vec{E}}^{(e)}}={{\vec{E}}^{(h1)}}+{{\vec{E}}^{(h2)}}+{{\vec{E}}^{(e)}},
\end{equation}
\begin{equation}\label{eq:4.3}
    \left\{ \begin{aligned}
  & E_{r}^{(h1)} \\ 
 & E_{z}^{(h1)} \\ 
\end{aligned} \right\}=\frac{ikR_{0}^{2}}{4\pi }\int\limits_{{{\Theta }_{1}}}^{{{\Theta }_{2}}}{d\theta '\int\limits_{0}^{2\pi }{d\varphi '}}\left\{ \begin{aligned}
  -\cos &\theta '\cos \varphi ' \\ 
 & \sin \theta ' \\ 
\end{aligned} \right\}\sin \theta '\frac{\exp (ik\tilde{R})}{{\tilde{R}}}H_{\varphi '}\left(R_0, \theta' \right),
\end{equation}
\begin{equation}\label{eq:4.4}
    \begin{aligned}
  & \left\{ \begin{aligned}
  & E_{r}^{(h2)} \\ 
 & E_{z}^{(h2)} \\ 
\end{aligned} \right\}=\frac{ikR_{0}^{2}R}{4\pi }\int\limits_{{{\Theta }_{1}}}^{{{\Theta }_{2}}}{d\theta '\int\limits_{0}^{2\pi }{d\varphi '}}\left\{ \begin{aligned}
  & R_0\sin \theta '\cos \varphi '-R\sin \theta  \\ 
 & R_0\cos \theta '-R\cos \theta  \\ 
\end{aligned} \right\} \\ 
 & \times \sin \theta '\left( \cos \theta \sin \theta '-\sin \theta \cos \theta '\cos \varphi ' \right)H_{\varphi '}\left(R_0, \theta' \right)\frac{\exp (ik\tilde{R})}{{{{\tilde{R}}}^{3}}}, \\ 
\end{aligned}
\end{equation}
\begin{equation}\label{eq:4.5}
    \left\{ \begin{aligned}
  & E_{r}^{(e)} \\ 
 & E_{z}^{(e)} \\ 
\end{aligned} \right\}=\frac{ikR_{0}^{2}}{4\pi }\int\limits_{{{\Theta }_{1}}}^{{{\Theta }_{2}}}{d\theta '\int\limits_{0}^{2\pi }{d\varphi '}\left\{ \begin{aligned}
  & \left( R_0\cos \theta '-R\cos \theta  \right)~\cos \varphi ' \\ 
 & -R_0\sin \theta '+R\sin \theta \cos \varphi ' \\ 
\end{aligned} \right\}}\sin \theta '\frac{\exp (ik\tilde{R})}{{{{\tilde{R}}}^{2}}}E_{\theta '}\left(R_0, \theta' \right), 
\end{equation}
\begin{equation}\label{eq:4.6}
    \tilde{R}=\sqrt{R_{0}^{2}+{{R}^{2}}-2RR_0\left( \cos \theta \cos \theta '+\sin \theta \sin \theta '\cos \varphi ' \right)}.
\end{equation}
Here $E_{\theta '}\left(R_0, \theta' \right)$, $H_{\varphi '}\left(R_0, \theta' \right)$ are the tangential components of the transmitted field on the surface of the ball determined by formulas~\eqref{eq:2.3}. The limits of integration in \eqref{eq:4.3} – \eqref{eq:4.5} are determined by the conditions that the aperture is a part of the object surface illuminated by CR which does not experience total internal reflection, in other words, $\left| {{\theta }_{i}} \right|<{{\theta }_{*}} \equiv \arcsin \left( {1}/{n} \right)$. 
One can show that   
${{\Theta }_{1}}=\max \left\{ {{\theta }_{p}}-{{\theta }_{*}}, \arcsin({a}/{R_0}) \right\}$, 
${{\Theta }_{2}}= \min\left\{ {{\theta }_{p}}+{{\theta }_{*}}, 2{{\theta }_{p}} \right\}$.

\section{\label{sec:5}Numerical results}

\begin{figure*}[t!]
    \centering
    \begin{subfigure}[t]{0.5\textwidth}
        \centering
        \includegraphics[width=\linewidth]{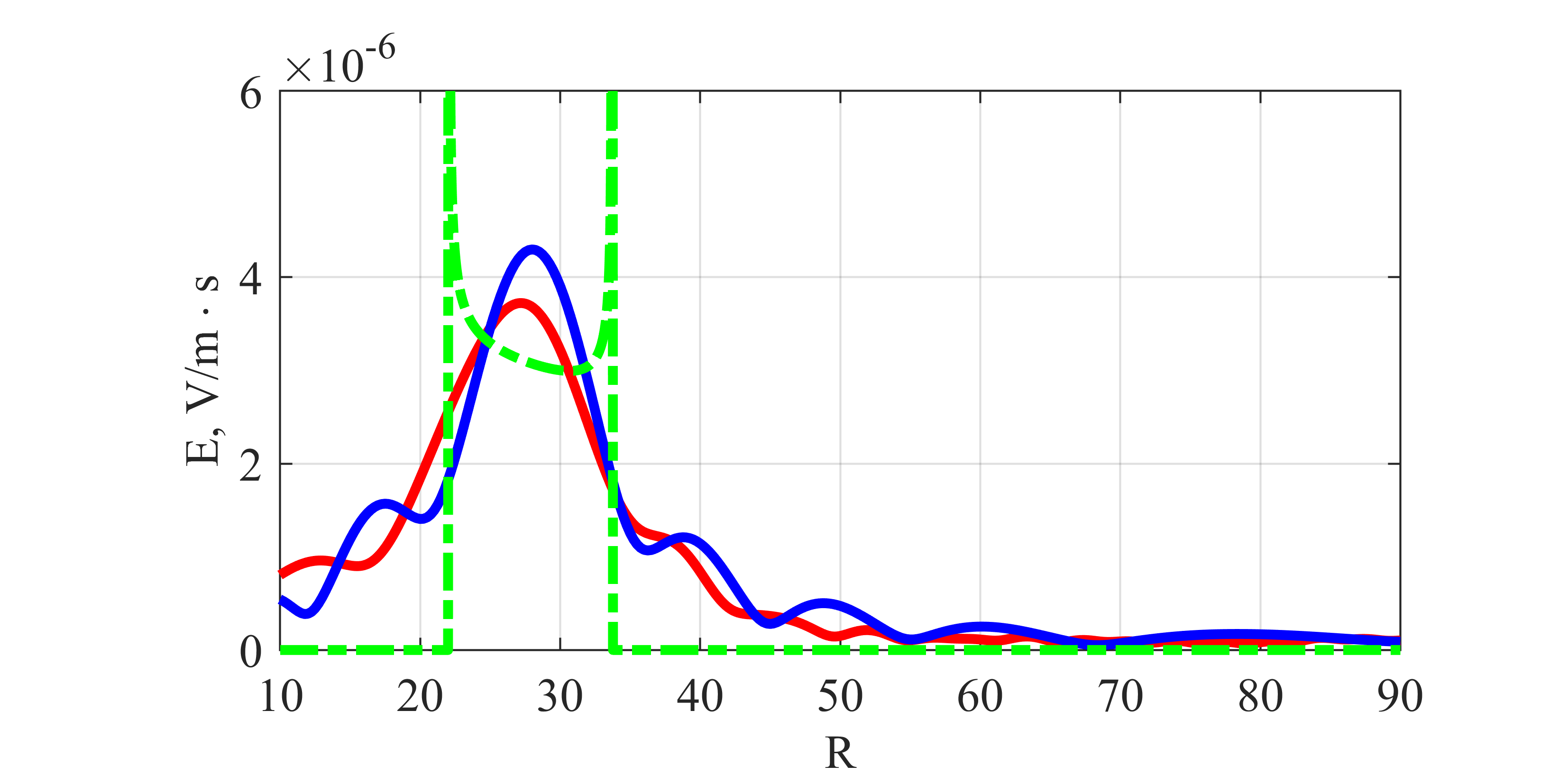}
        \caption{$\beta=0.8$, $R_0=30$;}\label{fig:3:a}
    \end{subfigure}%
    \begin{subfigure}[t]{0.5\textwidth}
        \centering
        \includegraphics[width=\linewidth]{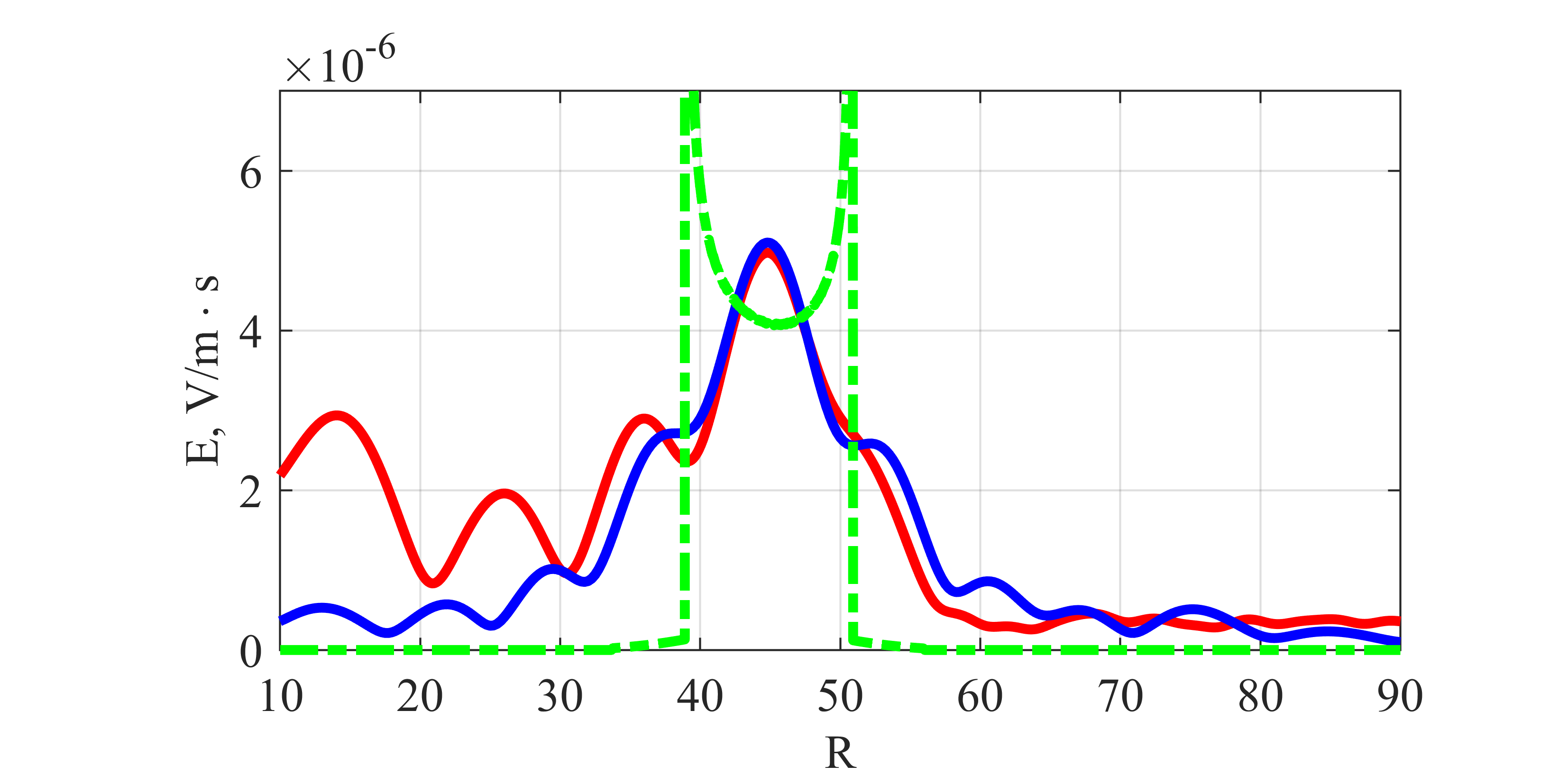}
        \caption{$\beta=0.999$, $R_0=30$;}\label{fig:3:b}
    \end{subfigure}
    \begin{subfigure}[t]{0.5\textwidth}
        \centering
        \includegraphics[width=\linewidth]{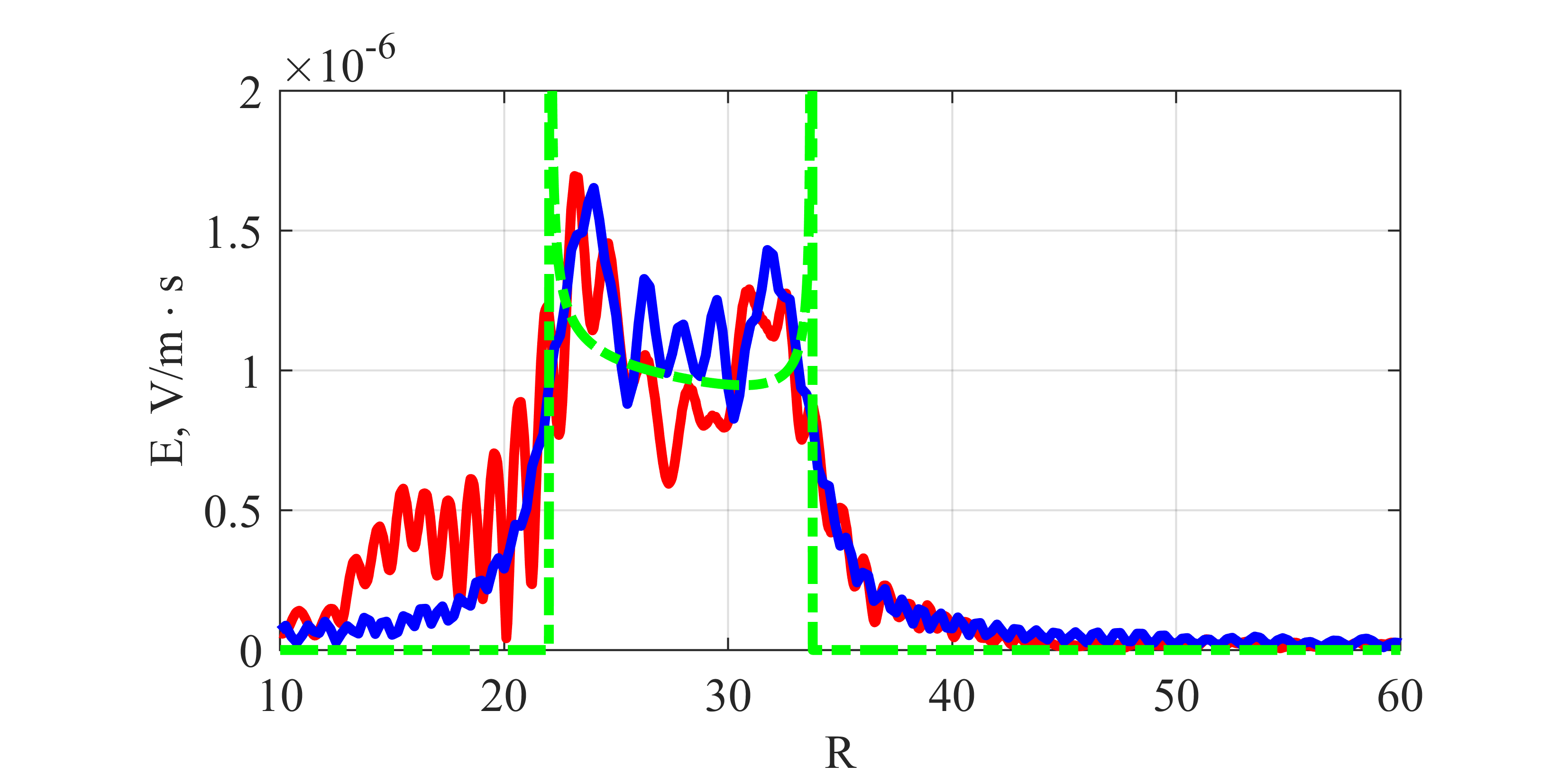}
        \caption{$\beta=0.8$, $R_0=300$;}\label{fig:3:c}
    \end{subfigure}%
    \begin{subfigure}[t]{0.5\textwidth}
        \centering
        \includegraphics[width=\linewidth]{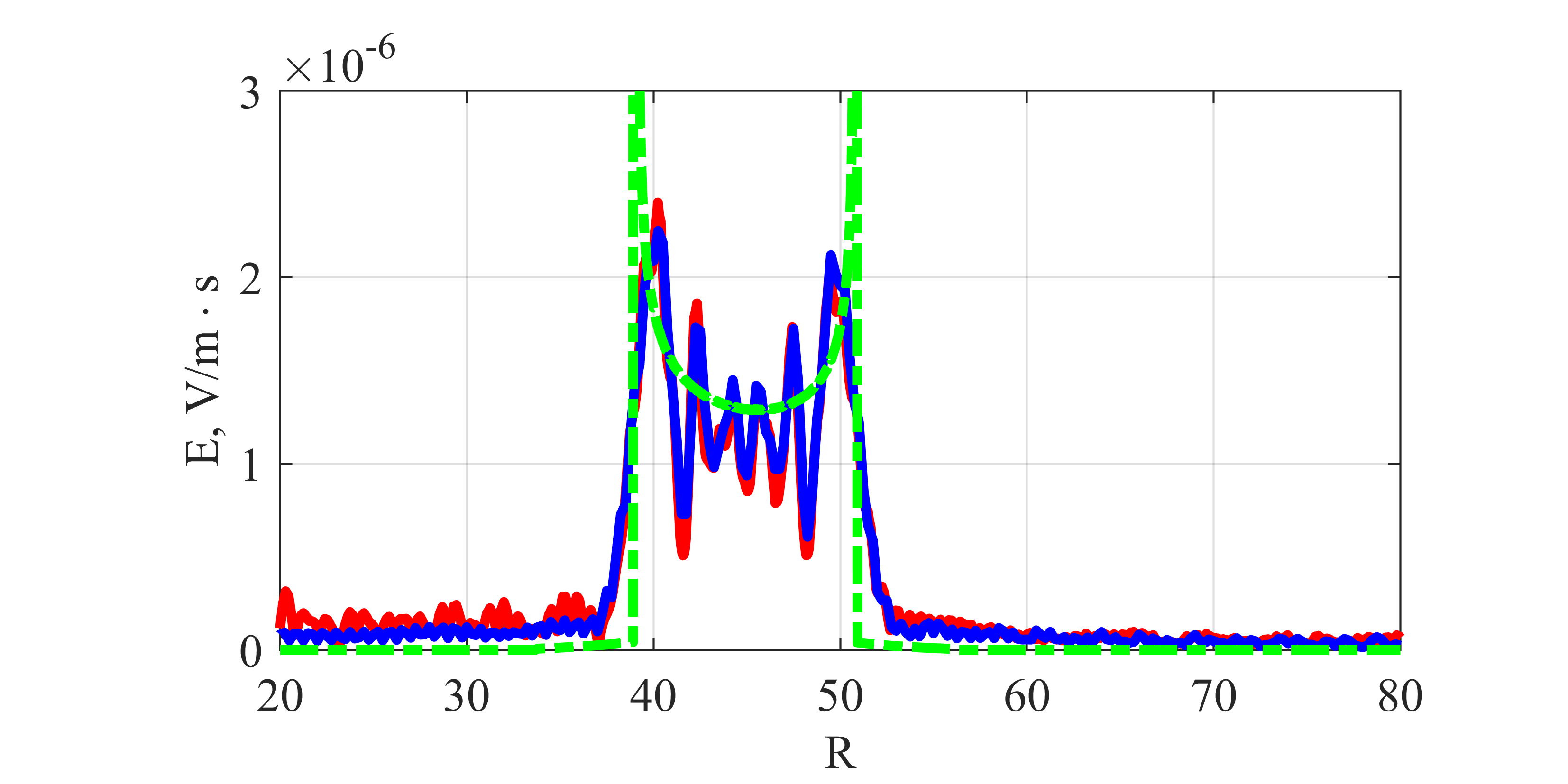}
        \caption{$\beta=0.999$, $R_0=300$;}\label{fig:3:d}
    \end{subfigure}
    \caption{Absolute value of Fourier-transform of electric field (${\text{V}}/{\text{m}}\cdot \text{s}$) depending on the angle $\theta $ (grad) for the following parameters: $q=1\,\text{nC}$, $a=c/\omega $, $\varepsilon =2$, $R=2{{R}_{0}}$; $\beta $ and ${{R}_{0}}$ (in ${c}/{\omega }$ units) are given under the plots. The curves were obtained on the basis of the ray optics method (dashed green curve) and the aperture one (solid blue curve). Red curves are the COMSOL simulations. }\label{fig:3}
\end{figure*}
\begin{figure*}[t!]
    \centering
    \begin{subfigure}[t]{0.5\textwidth}
        \centering
        \includegraphics[width=\linewidth]{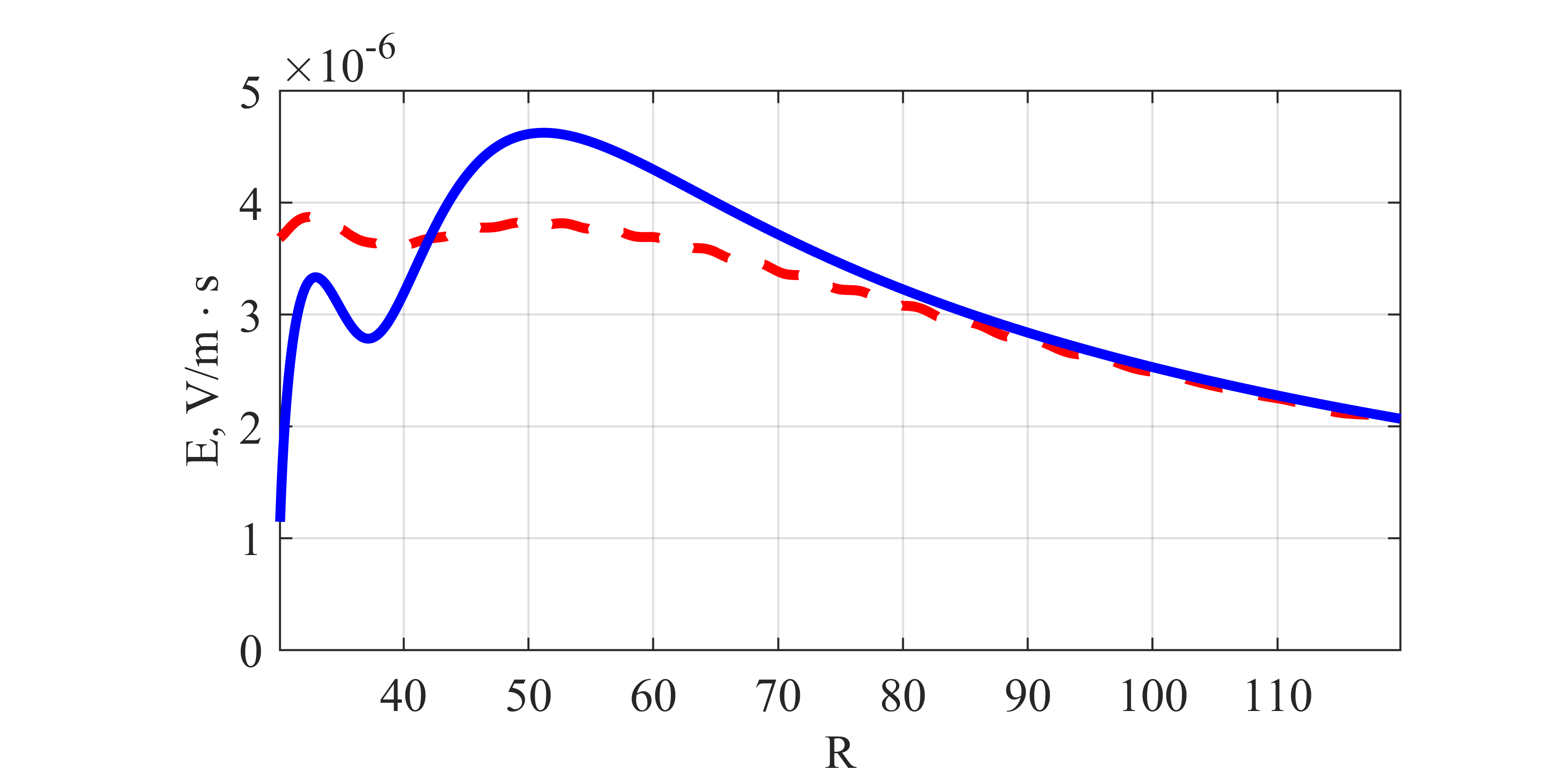}
        \caption{$\beta=0.8$, $R_0=30$, $\theta \approx 28\degree$;}\label{fig:4:a}
    \end{subfigure}%
    \begin{subfigure}[t]{0.5\textwidth}
        \centering
        \includegraphics[width=\linewidth]{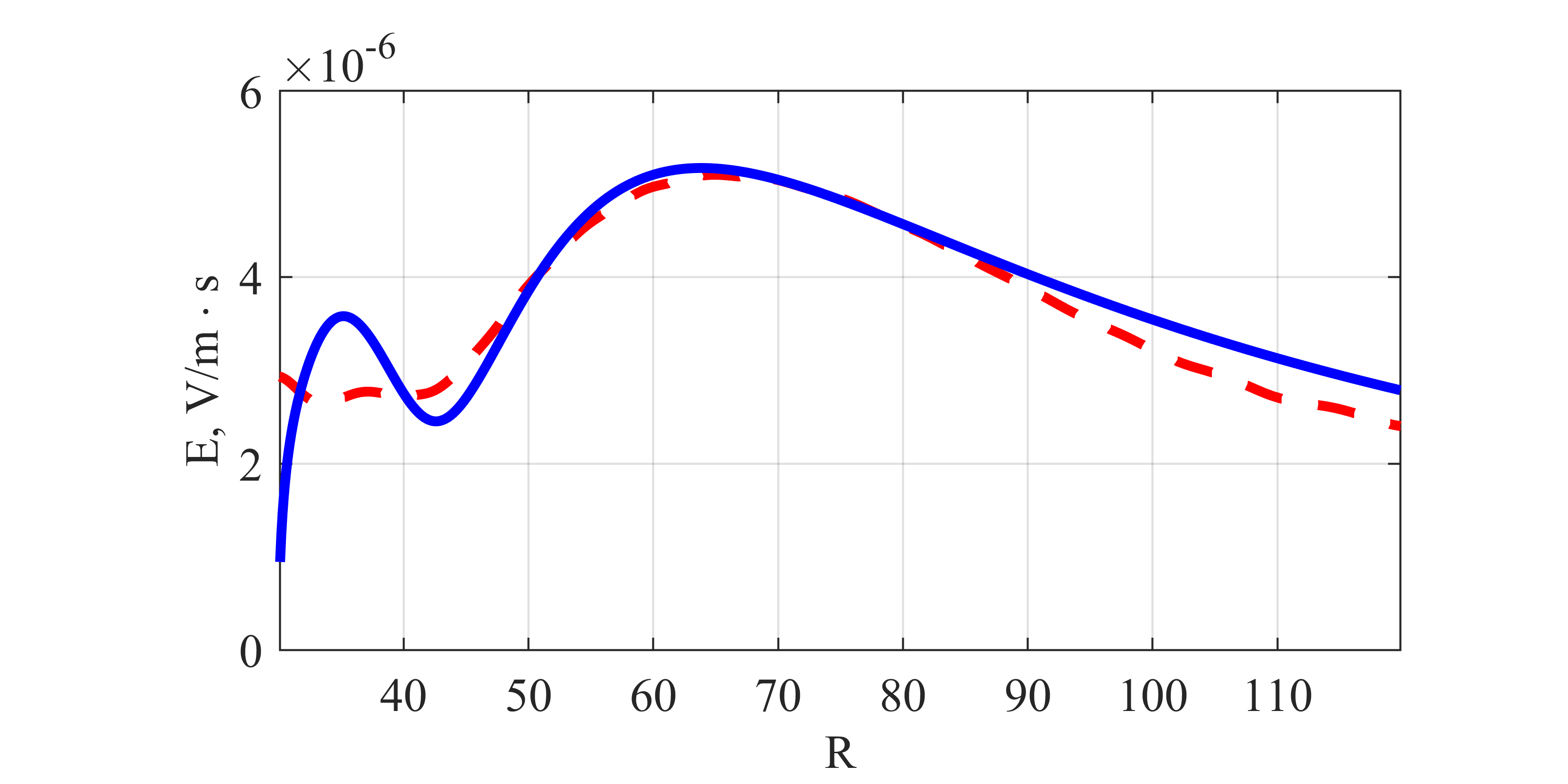}
        \caption{$\beta=0.999$, $R_0=30$, $\theta \approx 45\degree$;}\label{fig:4:b}
    \end{subfigure}
    \begin{subfigure}[t]{0.5\textwidth}
        \centering
        \includegraphics[width=\linewidth]{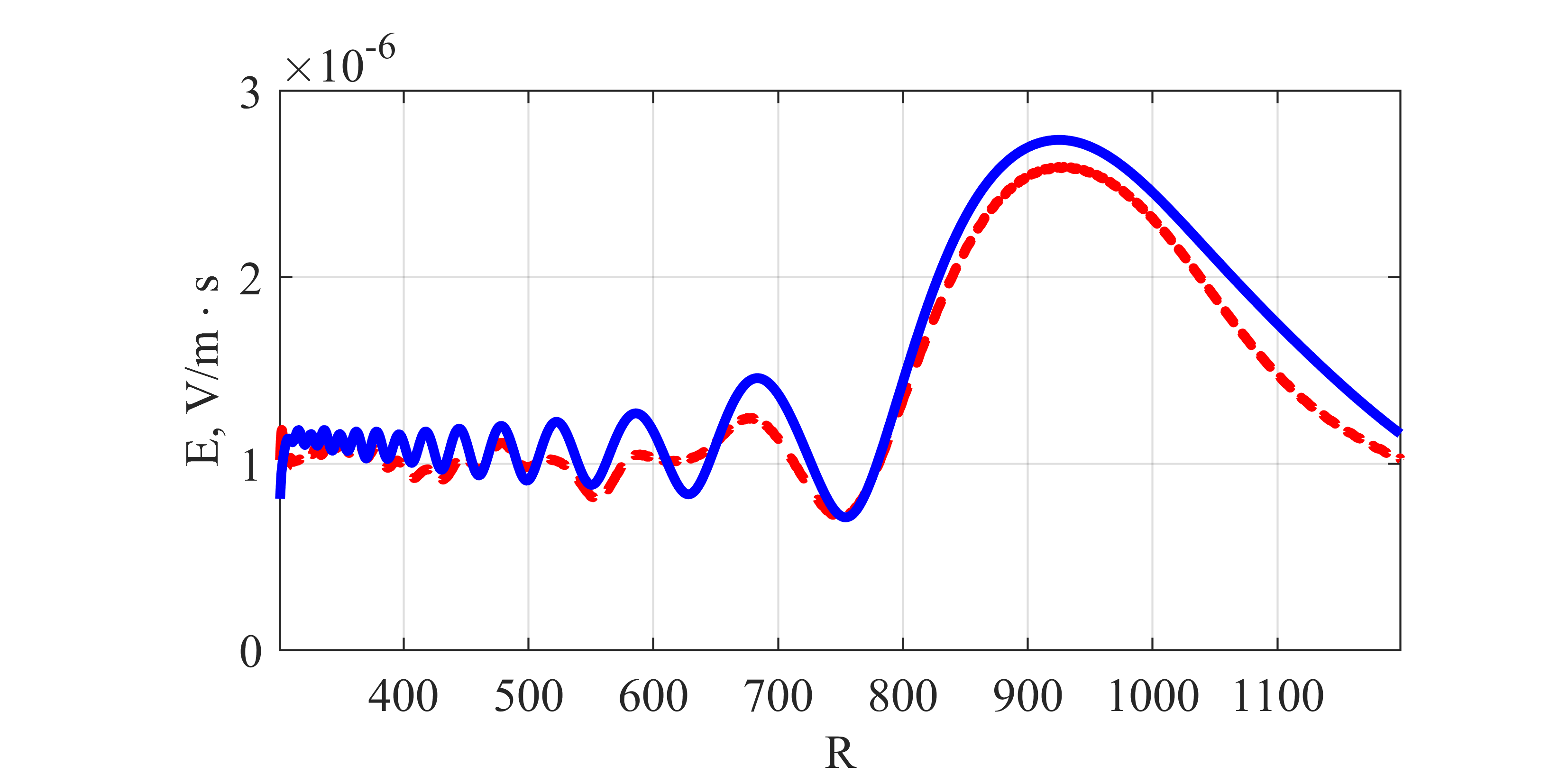}
        \caption{$\beta=0.8$, $R_0=300$, $\theta \approx 28\degree$;}\label{fig:4:c}
    \end{subfigure}%
    \begin{subfigure}[t]{0.5\textwidth}
        \centering
        \includegraphics[width=\linewidth]{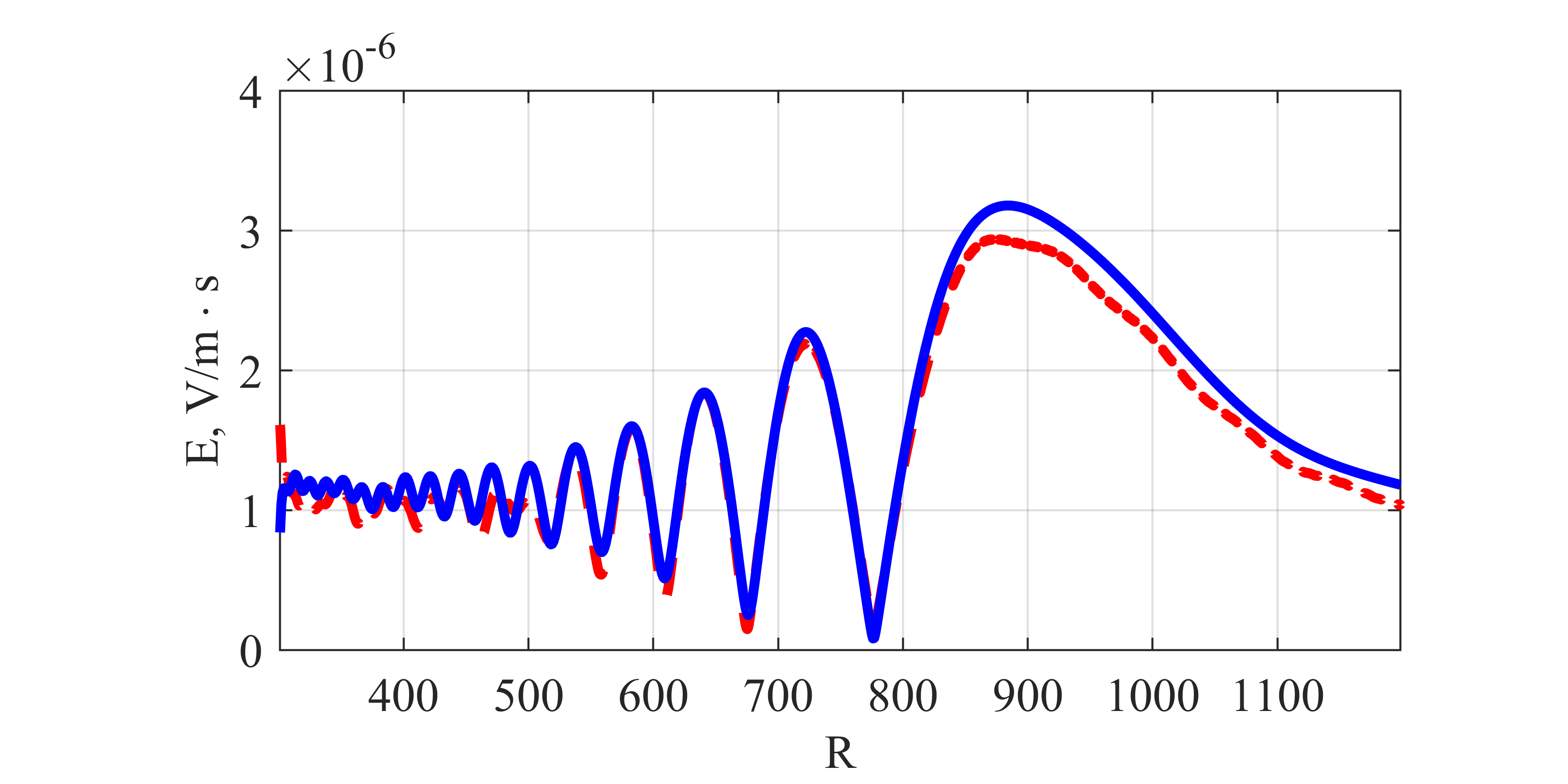}
        \caption{$\beta=0.999$, $R_0=300$, $\theta \approx 45\degree$;}\label{fig:4:d}
    \end{subfigure}
    \caption{Absolute value of Fourier-transform of electric field (${\text{V}}/{\text{m}}\;\cdot \text{s}$) depending on the distance $R$ (in ${c}/{\omega }$ units) for the angle $\theta ={{\theta }_{p}}$; $\beta $ and ${{R}_{0}}$ (in ${c}/{\omega }$ units) are given under the plots, other parameters are the same as in Fig.~\ref{fig:3}. Blue curves are obtained on the basis of the aperture method. Red curves are the COMSOL simulations. }\label{fig:4}
\end{figure*}
Figure~\ref{fig:3} shows the dependencies of the field magnitude on the angle $\theta $ obtained with the use of the ray optics (green curves) and aperture (blue curves) methods. The results of simulations performed with RF module of COMSOL Multiphysics (red curves) are demonstrated as well. We emphasize that all three methods give similar results even for not very large targets (the sphere radius is approximately  ${30}/{(2\pi )}\approx 5$ wavelengths in the Figures~\ref{fig:3:a} and~\ref{fig:3:b}). As we see, the ray optics results have typical (non-physical) gaps at the boundaries of the illuminated area. Naturally, the agreement between the results is better for the larger radius of the target (see Figures~\ref{fig:3:c} and~\ref{fig:3:d}, where sphere radius is approximately 50 wavelengths). It is natural as well that the aperture method gives the results which are closer to the COMSOL Multiphysics simulations in comparison with the ray optics. One can conclude that the ray optics method is suitable for estimation of the magnitude of the radiation field from the target with the radius of several wavelengths or larger. However, this technique is not suitable for studying the field behavior. The aperture technique is more general, gives more exact results and allows analyzing the field behaviour. 

Figure~\ref{fig:4} shows the dependencies of the field magnitude on the distance $R$ at the angle $\theta $ which is equal to the CR angle ${{\theta }_{p}}$. In this case, the refracted wave propagates normal to the sphere. It can be seen that with an increase in the distance from the sphere, the tendency toward an increase in the field is observed (complicated by the oscillations). The condensation of rays, which was noted within the ray optics examination (see Fig.~\ref{fig:2}), causes the observed phenomenon. The field starts to steadily decrease only after certain distance from the sphere surface. 

\section{\label{sec:6}Conclusion}

In this paper, two approaches have been applied for the analysis of radiation from a dielectric ball: the ray optics method and the aperture method. Each of them demonstrated a good coincidence with COMSOL Multiphysics (in the area of their applicability). However, the aperture method gave more exact results for the field structure. Numerous calculations performed for various parameters show that, as a rule, the error of the aperture method in the area of the largest magnitudes of the field is less than $10\%$ for the objects having the size of the order of 10 wavelengths. For larger objects, the error becomes even smaller. The main physical effects have been described. For example, it has been shown that, in the direction of the Cherenkov angle, the radiation field possesses an expressed maximum.

\section{Acknowledgments}

This research was supported by the Russian Science Foundation, Grant No. 18-72-10137. Numerical simulations with COMSOL Multiphysics have been performed in the Computer Center of the Saint Petersburg State University.

\end{document}